\documentclass[%
 reprint,
superscriptaddress,
 amsmath,amssymb,
 aps,
prb,
floatfix,
]{revtex4-2}

\usepackage{threeparttable}
\usepackage{setspace}
\usepackage{makecell}
\usepackage{amsmath}
\usepackage{graphicx}
\usepackage{dcolumn}
\usepackage{bm}
\usepackage{hyperref}
\usepackage[mathlines, pagewise]{lineno}
\usepackage{textcomp}
\usepackage{gensymb}
\usepackage{subfigure}
\usepackage{url}
\usepackage{xcolor}
\usepackage{hhline}
\usepackage[version=3]{mhchem} % Formula subscripts using \ce{}
\usepackage{multirow}

\begin{document}
%\linenumbers
%%% old titles for final choose:
%\title{Multifunctional GaON Nanolayer for Extending Functionality of GaN-based Devices}
%\title{Gallium oxynitride: A GaN-derived versatile nanolayer}
%\title{Gallium Oxynitride Nanolayer for Boosting GaN Surface Performance and Multifunction Extension}
%\title{GaN-derived epitaxial gallium oxynitride nanolayer for surface engineering}
%\title{Versatile Gallium Oxynitride Nanolayer Carved on GaN Surface}

\title{Lattice-aligned gallium oxynitride nanolayer for GaN surface enhancement and function extension}

\author{Junting Chen}\altaffiliation{Junting Chen and Junlei Zhao contributed equally to this work.}
\affiliation{Department of Electrical and Electronic Engineering, Southern University of Science and Technology, Shenzhen 518055, China}
\affiliation{Department of Electronic and Computer Engineering, The Hong Kong University of Science and Technology, Clear Water Bay, Kowloon, Hong Kong}

\author{Junlei Zhao}\altaffiliation{Junting Chen and Junlei Zhao contributed equally to this work.}
\affiliation{Department of Electrical and Electronic Engineering, Southern University of Science and Technology, Shenzhen 518055, China}

\author{Sirui Feng}
\affiliation{Department of Electronic and Computer Engineering, The Hong Kong University of Science and Technology, Clear Water Bay, Kowloon, Hong Kong}

\author{Li Zhang}
\affiliation{Department of Electronic and Computer Engineering, The Hong Kong University of Science and Technology, Clear Water Bay, Kowloon, Hong Kong}

\author{Yan Cheng}
\affiliation{Department of Electronic and Computer Engineering, The Hong Kong University of Science and Technology, Clear Water Bay, Kowloon, Hong Kong}

\author{Hang Liao}
\affiliation{Department of Electronic and Computer Engineering, The Hong Kong University of Science and Technology, Clear Water Bay, Kowloon, Hong Kong}

\author{Zheyang Zheng}
\affiliation{Department of Electronic and Computer Engineering, The Hong Kong University of Science and Technology, Clear Water Bay, Kowloon, Hong Kong}

\author{Xiaolong Chen}
\affiliation{Department of Electrical and Electronic Engineering, Southern University of Science and Technology, Shenzhen 518055, China}

\author{Zhen Gao}
\affiliation{Department of Electrical and Electronic Engineering, Southern University of Science and Technology, Shenzhen 518055, China}

\author{Kevin J. Chen}
\affiliation{Department of Electronic and Computer Engineering, The Hong Kong University of Science and Technology, Clear Water Bay, Kowloon, Hong Kong}

\author{Mengyuan Hua} \email{Corresponding author: huamy@sustech.edu.cn}
\affiliation{Department of Electrical and Electronic Engineering, Southern University of Science and Technology, Shenzhen 518055, China}

\begin{abstract}

Gallium nitride (GaN), as a promising alternative semiconductor to silicon, is of well-established use in photoelectronic and electronic technology. %such as solid-state lighting, wireless communication and power conversion. 
However, the vulnerable GaN surface has been a critical restriction that hinders the development of GaN-based devices, especially regarding device stability and reliability. 
Here, we overcome this challenge by converting the GaN surface into a gallium oxynitride (GaON) epitaxial nanolayer through an \textit{in-situ} two-step ``oxidation-reconfiguration”  process. 
The oxygen plasma treatment overcomes the chemical inertness of the GaN surface, and the sequential thermal annealing manipulates the kinetic-thermodynamic reaction pathways to create a metastable GaON nanolayer with  wurtzite lattice. 
This GaN-derived GaON nanolayer is a tailored structure for surface reinforcement and possesses several advantages, including wide bandgap, high thermodynamic stability, and large valence band offset with GaN substrate.
%This GaN-derived GaON nanolayer possesses several advantages, including wide bandgap, high thermodynamic stability, and large valence band offset with GaN substrate, making it a tailored layer for surface enhancement.
These enhanced physical properties can be further leveraged to enable GaN-based applications in new scenarios, such as complementary logic integrated circuits, photoelectrochemical water splitting, and ultraviolet photoelectric conversion, making GaON a versatile functionality extender.

\end{abstract}
\maketitle

%\begin{linenumbers}
\section{Introduction}

In parallel with conventional silicon-based devices, tremendous efforts have been directed towards exploring stable compound semiconductor materials and their applications over the past decades, including silicon carbide ~\cite{miao2020universal}, III-V compounds~\cite{delalamo2011nanometrescale, kuech2016growth}, and various oxides~\cite{wurdack2021ultrathin, wang2021ultrahigh}. 
Amongst a large multitude of candidates, GaN-based devices have gained wide acceptance as one of the successfully commercialized compound semiconductor systems, not only in well-established photoelectronic devices, power electronics, radiofrequency power amplifiers~\cite{chen2017gan, nela2021multi}, but also in emerging fields such as digital and quantum computing~\cite{zheng2021gallium, teo2021emerging}. 

Nevertheless, at its cutting edge, the advance of the GaN-based device encounters bottleneck restrictions due to the vulnerable GaN surface, especially for electronic devices. 
%In particular, 
The charging state of the GaN surface is easily changed by electric field through carrier trapping or detrapping, resulting in unstable electrical performance under switching operation, e.g., current collapse, threshold voltage shift, and worse block capability in GaN-based transistors~\cite{meneghini2021gan}. 
Furthermore, electric field tends to crowd near the surface in many electronic device structures, making it a weak point suffering hot electron bombardment and piezoelectric stress~\cite{SYang2021EDL}. 
Moreover, the exposed GaN surface is sensitive to the fabrication processes, exacerbating electric field non-uniformity and surface reliability issues. 
With GaN devices being scaled down toward emerging low-voltage applications, the surface effects are expected to be much more critical in the nano-scale devices featuring a large surface-to-volume ratio. 

A common remedy for these aforementioned issues is to exquisitely passivate the GaN surface utilizing deliberate chemical reactions. 
Although different passivation methods are still developing at a fast pace~\cite{chevtchenko2007study, huang2012effective}, the conventional approaches rely on depositing dielectric layers on pristine GaN surface. 
However, compared to external deposition, to achieve an uncontaminated passivation/GaN interface and insensitivity to the initial conditions of pre-passivated surface, an internal passivation layer derived intrinsically from GaN is more favorable. 

Oxides that can be formed by thermal oxidation (e.g., SiO$_{2}$ on Si~\cite{himpsel1988microscopic}) are relatively stable under air exposure and in aqueous environments by chemical nature, and their wide bandgaps in general are well suited for dielectric layer. 
However, one essential difficulty of passivating GaN surface by oxidation is the chemically inert features of GaN surface. 
Conventional thermal treatment below the decomposition temperature of GaN ($\sim$800 \degree C) is highly inefficient to form Ga$_{2}$O$_{3}$ intact nanolayers~\cite{schwartz1983analysis, delalamo2011nanometrescale, laukkanen2021passivation}.
Only mono- or bi-layer oxides can be formed with limited thicknesses of sub-1 nm~\cite{miao2010oxidation}. 
Instead of direct thermal oxidation, plasma-assisted process is a natural alternative approach to overcome the surface inertness of GaN with a higher reaction rate.
%because of a higher efficiency of oxidation.
Besides, another difficulty is that the fully oxidized Ga$_{2}$O$_{3}$ often adapts to the stable monoclinic $\beta$-phase with large lattice mismatch and different symmetry against hexagonal GaN (0001) substrate, resulting in high interface state density. 
Therefore, gallium oxynitride (GaON) nanolayer formed by partial oxidation of GaN is an alternative material of passivation, because of its better compatibility with GaN substrate.

Despite the promising features of plasma-assisted partial oxidation as a surface passivation of GaN, investigation on achieving a metastable GaON epitaxial nanolayer on GaN is still at an immature stage, and the underlying formation mechanisms need a in-depth understanding.
%GaN assisted by surface plasma treatment to form a gallium oxynitride (GaON) nanolayer is a promising way of passivation, because of a higher efficiency of oxidation and yet a better compatibility with GaN substrate. 
%Nevertheless, the investigation on the plasma-assisted oxidation approach to achieve a metastable GaON epitaxial nanolayer on GaN is still at an immature stage, and the underlying formation mechanisms need a in-depth understanding.
In this work, we will show a simple and robust oxidation-reconfiguration approach to achieve well-controlled near-surface ($<$ 5 nm) partial oxidation of GaN to enhance the surface and, moreover, lead to a versatile epitaxial nanolayer (GaON) on GaN surface. 
The GaON nanolayer is formed by a two-step oxidation process, using controllable remote oxygen plasma and sequential thermal annealing. 
Leveraging well-established experimental and computational methods, the formation mechanism and detailed structural and electrical characteristics of the GaON nanolayers are investigated. 
Control of the thickness, composition, and phase separation is readily accomplished by tuning the kinetic pathways of the intentional GaN-surface oxidation reaction. 
Finally, based on these findings of the nanosized GaON-on-GaN lamellar system, we demonstrate and discuss the potential applications in GaN-based power devices~\cite{zhang2021characterization}, GaN $p$-channel field effect transistors ($p$-FETs)~\cite{zheng2021gallium}, photoelectron-chemical (PEC) water splitting~\cite{phivilay2013fundamental, ma2019low}, and ultraviolet photodiodes~\cite{su2020low, Song2021Self}. 
This contribution can open up new prospects for future study on the GaON nanolayer as a versatile platform retaining intrinsic compatibility with a variety of GaN-based cutting-edge applications. 

\section{Results and Discussion}

\subsection{Two-Step formation of GaON nanolayer}

\begin{figure*}[ht!]
  \centering
  \includegraphics[width=15cm]{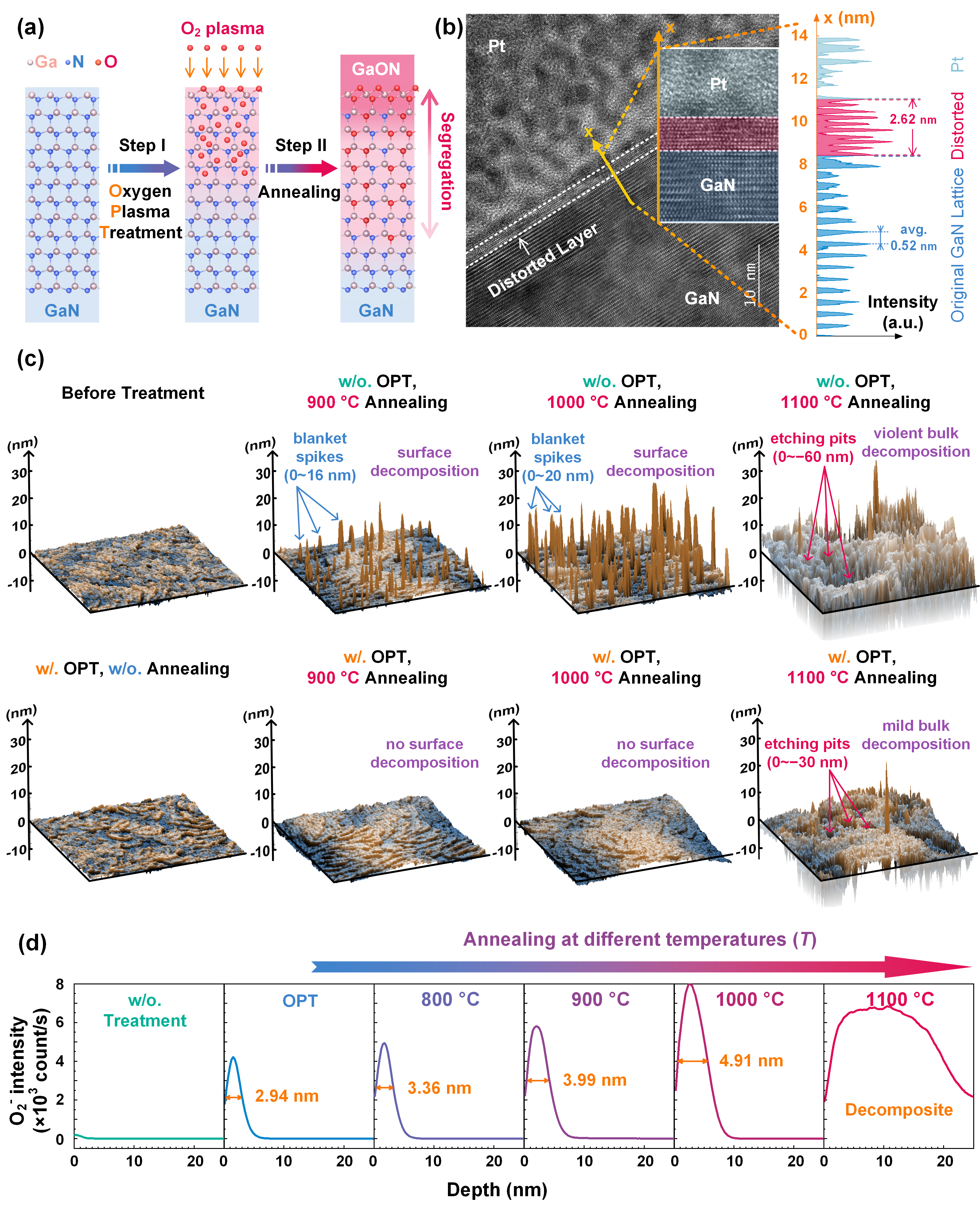}
  \caption{(a) The schematics of the two-step formation of GaON nanolayer on the GaN(0001) surface. (b) TEM image of the sample after OPT and 900 \degree C annealing, and the line profile of the pixel intensity along $x$ direction indicating the two distinct layers with the different Ga-plane distances. (c) AFM-measured surface profiles of the samples after different treatments. The blanket spikes and etching pits are seen in the surface/bulk-decomposed samples. The scanning area of AFM is 5 $\times$ 5 $\mu m^{2}$. (d) Secondary O$_{2}^{-}$ ion intensity (used to indicate O concentration in the GaON nanolayers) from ToF-SIMS to analyze the samples after different treatments. More AFM images and full profiles of ToF-SIMS are shown in the Supplementary Information Figure S1-S2.}
  \label{fig:Fig1}
\end{figure*}

The two-step "oxidation-reconfiguration" process of the GaON nanolayer involves oxygen plasma treatment (OPT) followed by sequential high-temperature annealing in an N$_{2}$ atmosphere, as schematically illustrated in Figure~\ref{fig:Fig1}a. 
More detailed conditions of fabrication can be found in the Experimental Section. 
Compared with direct thermal oxidation, OPT can assist oxygen atoms in penetrating into GaN for several nanometers and overcome the chemical inertness of GaN in the ambient atmosphere. 
However, this far-from-equilibrium process can cause energetically unfavorable defect species, such as Frenkel pairs and extra O interstitials, to appear in the GaN lattice transiently. 
Therefore, high-temperature annealing is conducted subsequently to promote the defect healing process, where the O atoms will substitute the N atoms to form stable O$_\mathrm{N}$ sites~\cite{wright2005oxygen}, and release the additional N atoms to the gas phase. 
Moreover, rather unexpectedly, a segregation of the O composition is observed during the annealing, resulting in two distinct high- and low-oxygen-concentration layers, as illustrated in Figure~\ref{fig:Fig1}a (after annealing). 
This aspect will be elucidated by more experimental characterizations later in this section. 

One major advantage owing to oxygen incorporation is the enhanced thermal stability of the GaN surface. 
The bond formed between Ga and O is more energetically favorable and thermodynamically stable than that between Ga and N, since O has a lower electron affinity than N. 
Thus, better thermal stability of the GaON nanophase is expected, which is verified by the surface morphology after thermal annealing (Figure~\ref{fig:Fig1}c). 
After OPT, four groups of samples were annealed at 800/900/1000/1100 \degree C separately, and one group remained without annealing for reference. 
Another set of pristine GaN samples was only annealed but without OPT to verify the enhanced thermal stability of the GaON nanolayer. 
The initial surface of the GaN sample (before treatment) is of high quality with a small surface roughness ($R_{q}$) of 0.588 nm.
Its comparison with the OPT-processed sample (w/. OPT, w/o.  annealing) confirms that OPT alone has no significant effect on the surface morphology, as clear atomic steps can be found in both samples. 
However, the further comparison between the annealed samples shows clearly different trends: for the samples without OPT, the samples start to decompose when the annealing temperatures reach 900 \degree C (w/o. OPT, 900 \degree C annealing) due to the facet-selected thermal decomposition of the GaN surface~\cite{koleske2001gan, choi2002surface}. 
The ``blanket spikes” appearing on the surface are the remaining Ga (oxidized in ambient conditions later) after the decomposition of GaN. 
As a comparison, the samples with OPT resist decomposition up to 1000 \degree C (w/. OPT, 1000 \degree C annealing). 
At 1100 \degree C, both samples with and without OPT have uniformly speared hexagonal ``etching pits” (small caves appearing at the surface), which indicate the bulk decomposition starting from the threading dislocation region ~\cite{hino2000characterization}. 
The overall AFM images of all the samples can be found in Supplementary Information Figure S1.

A representative transmission electron microscopy (TEM) image of the GaON/GaN stack is shown in Figure~\ref{fig:Fig1}b, which is obtained on the sample with OPT and 900 \degree C annealing. 
According to the line profile (right panel in Figure~\ref{fig:Fig1}b), the average distance between two Ga planes (where the peak intensity appears) in the bulk GaN region ($\sim$0.52 nm) is in good agreement with the lattice constant $c$ ($\sim$0.518 nm) of wurtzite GaN. 
However, noticeable lattice distortion can be seen near the surface, where the average distance between two Ga planes shrinks to 0.29 nm, and the total thickness of the distorted layer is 2.62 nm. 
A sharp boundary can be found between the distorted layer and the undistorted GaN lattice underneath.

Moreover, time-of-flight secondary-ion mass spectroscopy (ToF-SIMS) is adopted to probe the local composition of the near-surface region, as shown in Figure~\ref{fig:Fig1}d. 
The thickness of the GaON nanolayer is determined by the full width at half maximum (FWHM) of the normalized intensity of the secondary O$_{2}^{-}$ ion profiles. 
The ToF-SIMS measurements reveal that the OPT alone leads to an oxidized layer with a thickness of 2.94 nm, and the thickness extends to 3.36/3.99/4.91 nm after annealing at 800/900/1000 \degree C, respectively. 
Therefore, the thicknesses of oxidized nanolayers are well controlled in the nm range and monotonically increase with the annealing temperatures. 
We note that the thickness measured from ToF-SIMS profiles is unambiguously larger than the one of the distorted layers seen in the TEM image (3.99 nm versus 2.62 nm). 
This result indicates that a part of the oxide layer under the distorted layer can maintain the original wurtzite lattice, and hence not be identified in the TEM image, which will be further supported by the DFT calculation in the Section 2.2. 
The full profiles of ToF-SIMS are shown in the Supplementary Information Figure S2.

\begin{figure*}[ht!]
  \centering
  \includegraphics[width=17cm]{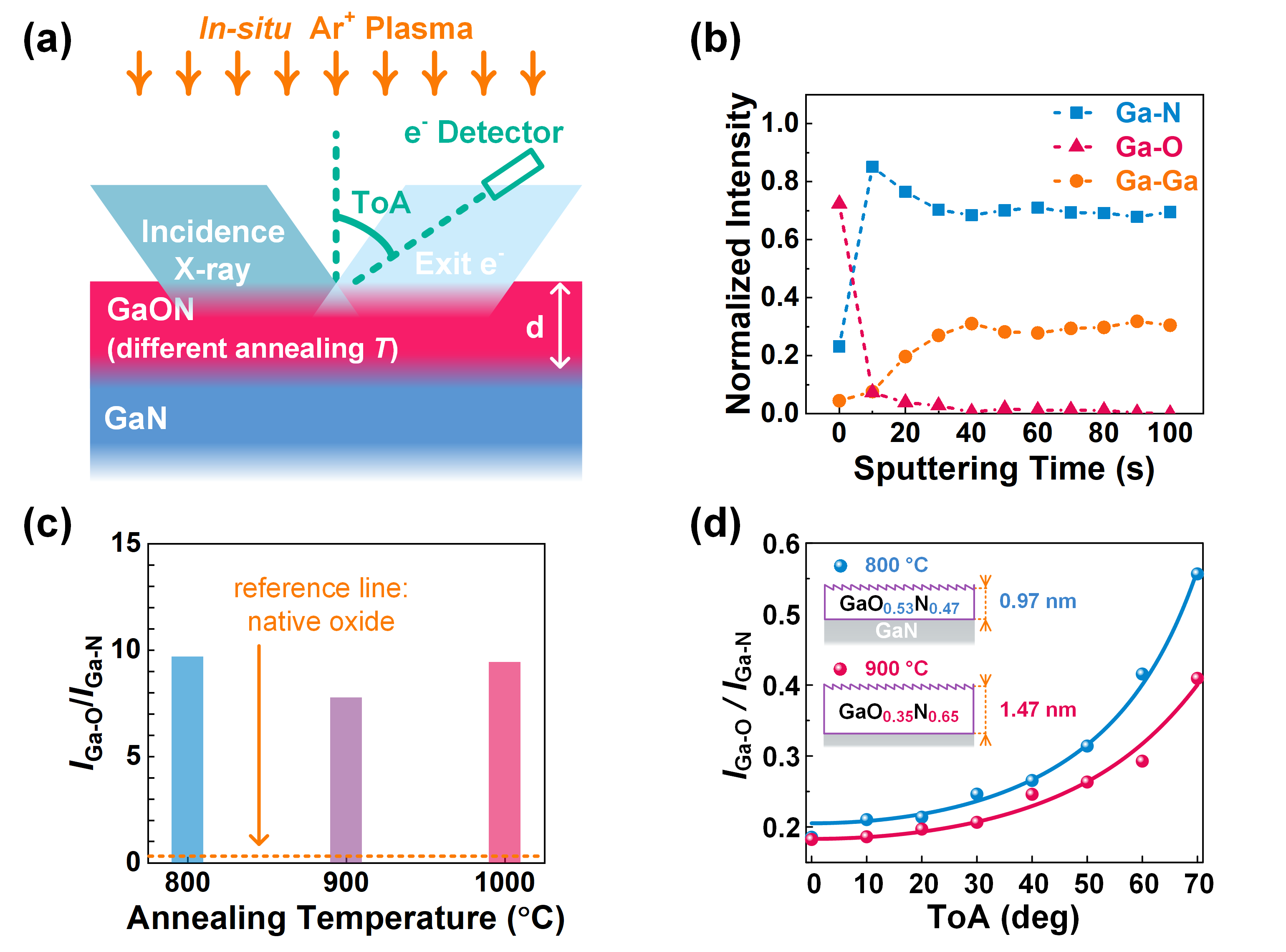}
  \caption{(a) Illustration of XPS measurement setup applied in this work. The sample that being used (with OPT and different annealing time), the duration of \textit{in-situ} Ar$^{+}$ sputtering, and the take-off angle (ToA) are the $x$-axis parameters in the following plots (b-d). 
  (b) The XPS peak intensities of Ga-N, Ga-O, Ga-Ga bonds at different \textit{in-situ} Ar$^{+}$ sputtering time. 
  The measured sample is with OPT and 900 \degree C annealing. 
  The curves are normalized to the total intensity of the corresponding Ga-3$d$ core-level peaks.
  (c) The intensity ratios between Ga-O and Ga-N bond at the surface of the different samples (with OPT) are not significantly affected by the annealing temperatures. 
  (d) The intensity ratio between Ga-O and Ga-N bond at different ToA of samples after OPT, 800/900 \degree C annealing, and 20 second of \textit{in-situ} Ar$^{+}$ sputtering. 
  The inset figures illustrate the fitted parameters from Equation~\ref{eq:Eq1}.
  More detailed analyses of the XPS data are shown in Supplementary Information Figure S3, S4 and S5.}
  \label{fig:Fig2}
\end{figure*}

The degree of oxidation is further analyzed by X-ray photoelectron spectroscopy (XPS). 
To understand the bonding conditions of the GaON nanolayers, XPS measurements are conducted on different samples with varying durations of \textit{in-situ} Ar$^{+}$ sputtering and the take-off angles (ToA), as illustrated in Figure~\ref{fig:Fig2}a. 
The spectra of the Ga-3$d$ core level are used to analyze the local chemical bonding environments. 
The area ratios of the Ga-N, Ga-O, and Ga-Ga subpeaks to the corresponding total Ga-3$d$ peaks (these ratios are referred to as the normalized intensity, \textit{I}) are used to determine the composition of the nanolayer.  

The sample with OPT and 900 \degree C annealing was first measured after every 10-second stepwise \textit{in-situ} Ar$^{+}$ surface sputtering, as shown in Figure~\ref{fig:Fig2}b. 
Initially, the non-sputtered (at 0 s) nanolayer has a very high concentration of oxygen. 
After the first 10-s sputtering, the fraction of Ga-O bonds decreases drastically, whereas the fraction of Ga-N bonds increases rapidly, revealing a clear shift of the surface chemical bonding environment. 
From 10 to 30 s, the fraction of Ga-O bonds evolves slowly towards zero. 
This change in the slope of the Ga-O curve, together with the TEM image and the SIMS data, strongly evince the existence of a near-surface high-oxygen-concentration (HOC) nanolayer and an underneath low-oxygen-concentration (LOC) nanolayer. 
After 40 s, the fraction of Ga-O bonds keeps around a negligible value, suggesting that the GaON nanolayer has been completely sputtered. The rise of the Ga-Ga bond intensity is because of the preferential sputtering of the light atoms (O and N), leaving a thin layer of metallic Ga on the surface~\cite{huang2018angular, lewandkow2021interface}.

The Ga-O/Ga-N bond ratio of the HOC-GaON layer can be directly obtained from the surface XPS profiles without sputtering, which is 9.68/7.77/9.44 after 800/900/1000 \degree C annealing, respectively (Figure~\ref{fig:Fig2}c). 
To obtain the XPS profiles of the underneath LOC-GaON nanolayers, the oxygen-rich layer needs to be removed by Ar$^{+}$ sputtering for 20 s \textit{in-situ} before the XPS measurements. 
Both the O/N atomic ratio ($x$/$y$) and the thickness of the remaining LOC-GaON layer can be obtained from the angle-resolved XPS (ARXPS) data~\cite{Seah2004ARXPS}. 
The Ga-O/Ga-N intensity ratios ($I_\mathrm{Ga-O}/I_\mathrm{Ga-N}$) in the Ga-3$d$ core level are plotted against the take-off angle (ToA), as shown in Figure~\ref{fig:Fig2}d. 
The $I_\mathrm{Ga-O}$ is contributed by the LOC-GaON nanolayer, while the $I_\mathrm{Ga-N}$ is from both the GaON nanolayer and the GaN substrate beneath. 
The $x$/$y$ can be quantitatively extracted from $I_\mathrm{Ga-O}/I_\mathrm{Ga-N}$ with the formula:
\begin{equation} \label{eq:Eq1}
    \frac{ I_\mathrm{Ga-O} }{ I_\mathrm{Ga-N} } = 
    \frac{ 1-\exp{[-d/\lambda\cos{(\theta)}]} }{y/x+\exp{[-d/\lambda\cos{(\theta)}]}},
\end{equation}
where $\lambda$ is the electron inelastic mean free path of 2.5 nm in GaN~\cite{nist2010imfp}, $d$ is the thickness of the remaining LOC-GaON nanolayer after sputtering, and $\theta$ is the ToA. 
For the 800 \degree C-annealed sample, the extracted $d$ is 0.97$\pm$0.08 nm, and $x$/$y$ is 1.15/1; and for the 900 \degree C-annealed sample, the extracted $d$ is 1.47$\pm$0.14 nm, and $x$/$y$ is 0.54/1. 
The thickness of GaON nanolayer in the 900-\degree C-annealed sample is 0.50$\pm$0.22 nm thicker than that of the 800 \degree C-annealed sample, which is well consistent with the difference in the corresponding SIMS data (0.63 nm in Figure~\ref{fig:Fig1}c). 
The more details of the XPS spectra are shown in Supplementary Information Figure S3, S4 and S5.

It is worth mentioning that this oxidation process is highly repetitive and tunable. 
The GaON nanolayer with various surface oxygen concentrations and thickness can be obtained as summarized in Table~\ref{Table1}, demonstrating its flexibility and wide application potential on the GaN-based device platform. 
The control of surface composition can be achieved by the total oxygen flux (controlled with time duration in Step I)~\cite{LZhang2021EDL}, and the thickness of the GaON nanolayer can be tuned with annealing temperature in Step II. 

\begin{table*}[ht!] 
\caption{(Al)GaON nanolayer originated from different fabrication parameters. The coil power and platen power are used to generate and control the oxygen plasma in the ICP chamber.}
\centering
  \begin{tabular}{@{}c|ccc|cc|cc|c@{}}
    \hline
    \multirow{3}{4em}{Material} & \multicolumn{3}{|c|}{\textbf{Step I: OPT}} & \multicolumn{2}{|c|}{\textbf{Step II: Annealing}} & \multicolumn{2}{|c|}{\textbf{Result}} & \multirow{3}{2em}{$\left[\textrm{Ref.}\right]$} \\
    \cline{2-8}{}
            & Coil          & Platen    & Duration  & Temp.        & \multirow{2}{4em}{Ambience}  & Surface   & Oxide      &                       \\
            & Power         & Power     & (min)     & (\degree C)  &                              & Ga-O/Ga-N & Depth      &                       \\
            & (W)           & (W)       &           &              &                              &      & (nm)       &                       \\
    \hline
    $p$-GaN & 50            & 20        & 5         & 800          & N$_{2}$                      & 9.68      & 3.36       & This work             \\
    $p$-GaN & 50            & 20        & 5         & 900          & N$_{2}$                      & 7.77      & 3.99       & This work             \\
    $p$-GaN & 50            & 20        & 5         & 1000         & N$_{2}$                      & 9.44      & 4.91       & This work             \\
    $p$-GaN & 50            & 30        & 4         & 800          & N$_{2}$                      & 5.40      & 4.30       & ~\cite{LZhang2021EDL} \\
    GaN     & 10            & 10        & 10        & 780          & NH$_{3}$                     & 0.95      & unknown    & ~\cite{MHua2018IEDM}  \\
    AlGaN   & 35            & 15        & 6         & 780          & NH$_{3}$                     & 0.30      & unknown    & ~\cite{SYang2021EDL}  \\
    \hline
  \end{tabular}
  \label{Table1}
\end{table*}

\subsection{Formation mechanism and band structure of GaON nanolayer}

So far, we have systematically investigated the structural and compositional properties of the GaON nanolayer using different experimental methods and analyses. 
However, the exact driving force behind the formations of the initial plasma-oxidized nanolayer (Step I) and later the segregated HOC- and LOC-GaON nanolayers (Step II) remain unclear. 
We will elucidate the formation of the HOC- and LOC-GaON nanolayers with computational methods, but before that, how the oxygen plasma can overcome the chemical inertness of the GaN (0001) surface is discussed.

In principle, the initial plasma-assisted penetration depth of oxygen involves the codependency between the plasma and the substrate. 
The estimated average kinetic energy of oxygen ion in inductively coupled plasma (ICP) chamber is about $\sim$10 eV which is about two orders of magnitude larger than that of a thermally heated oxygen gas (e.g. 0.165 eV/atom at 1000 \degree C).
Therefore, the bond breaking caused by oxygen plasma is not only electron charge transfer through surface adsorption, which can be solely understood by chemical reaction theory, but rather a localized thermal spike and lattice distortion caused by inelastic ion collision within the threshold energy of surface sputtering. 
As a result, in our experiments Step I, the oxygen plasma create a highly active condition with negligible reaction energy barriers for oxidation, so the O atom will penetrate into GaN surface a few nanometers and stay in non-equilibrium transient configurations.  

\begin{figure*}[ht!]
  \includegraphics[width=17cm]{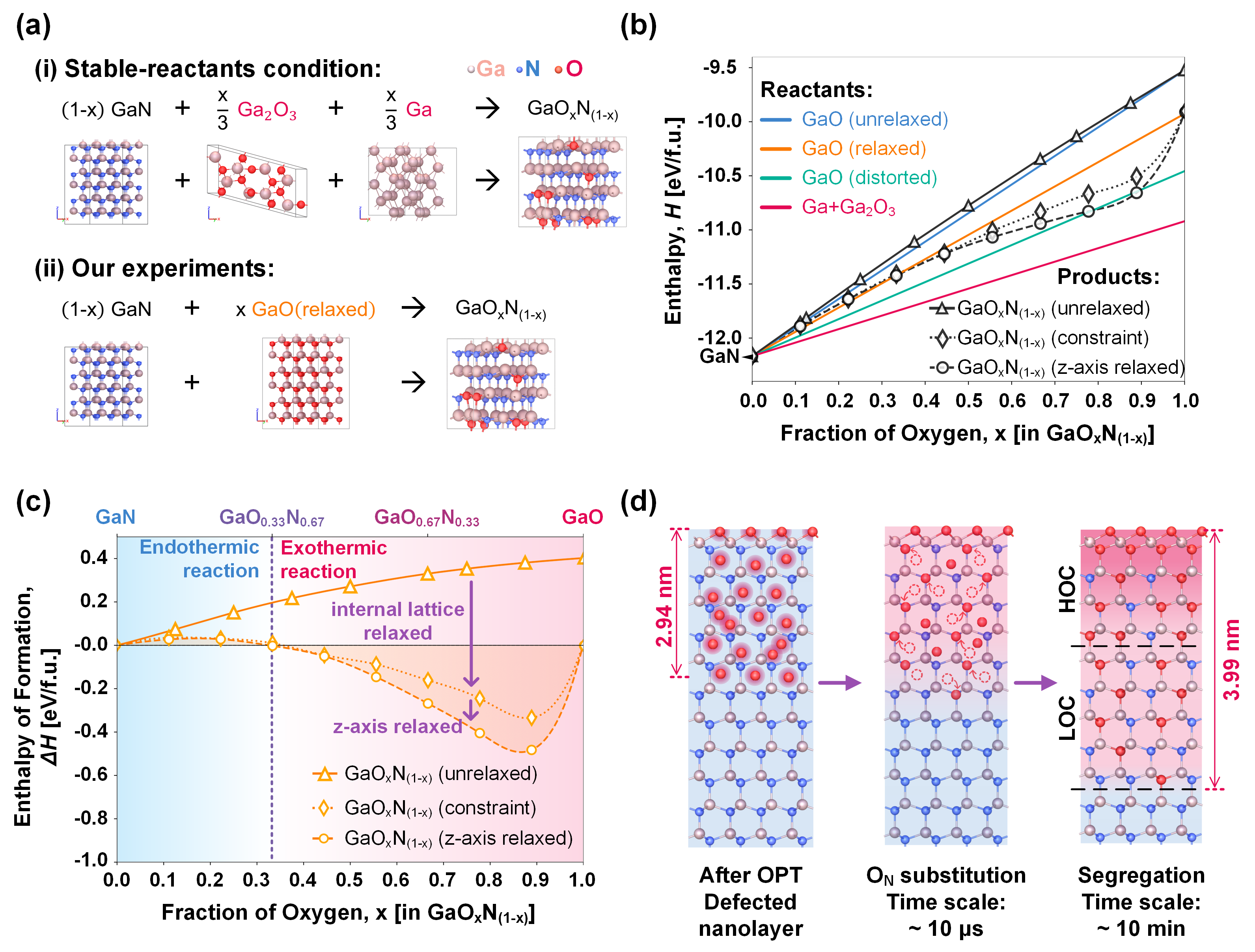}
  \caption{(a) Chemical reaction equations with conditions: (i) stable reactants and (ii) in our experiments. 
  (b) Enthalpy, $H$, of the different GaON states. The colored lines are the linear combinations of the reactants, i.e., GaN combined with three different GaO (and Ga+Ga$_{2}$O$_{3}$) states. The black curves are calculated using DFT. 
  %(b) Referenced reactants at different states. 
  %The colors of the labels correspond to the ones of the reference lines in (a). 
  (c) Enthalpy of formation (heat of mixing), $\Delta H$, against the reference line of the reactants ``GaN + GaO (relaxed)" as the orange line shown in (b). 
  (d) A schematic illustration of the fast local oxygen substitution and the following segregation of the HOC and LOC nanolayers. 
  The exemplary GaO$_\mathrm{0.33}$N$_\mathrm{0.67}$ configurations calculated by DFT are shown in Supplementary Information Figure S6.
  }
  \label{fig:Fig3}
\end{figure*}  

Thermodynamically, it is clear that the formation of the local substitutional O$_\mathrm{N}$ site is energetically favorable~\cite{wright2005oxygen}.
However, the overall supercell mapping through a wide range of oxygen composition is needed to understand the formation of the HOC- and LOC-GaON nanolayers, especially in the epitaxial nanolayer where the atomic strain induced by the GaN substrate can have a significant constraint effect on the reaction energy landscape. 
Therefore, to shed light on the formation and segregation mechanisms of the HOC- and LOC nanolayers, we invoke the \textit{ab-initio} model at the density functional theory (DFT) level to calculate the enthalpy ($H$) of the solid-state reactions under different conditions.
%, as shown in Figure~\ref{fig:Fig3}a. 

On the one hand, 
considering the solid-state chemical reaction equation with the stable reactants as shown in Figure~\ref{fig:Fig3}a (i)
%\begin{equation} \label{eq:Eq2}
%    (1-x) \mathrm{GaN} + \frac{x}{3} \mathrm{Ga}_{2}\mathrm{O}_{3} + \frac{x}{3} \mathrm{Ga} \Longrightarrow %\mathrm{GaO}_{x}\mathrm{N}_{(1-x)},
%\end{equation}
, where the reactants on the left-hand side 
are GaN in the wurtzite phase, Ga$_{2}$O$_{3}$ in $\beta$-phase and Ga in the orthorhombic phase, and the product on the right-hand side 
is a neutrally charged GaO$_{x}$N$_{(1-x)}$ supercell with the $x$ fraction of the N sites randomly substituted by O atoms at different strain states (these states will be explained later). 
%The exemplary GaO$_{x}$N$_{(1-x)}$ supercells can be found in Supplementary Information Figure S6.
The solid-state reaction in our experiments (OPT and annealing) retains mainly the wurtzite GaN lattice, and the reaction proceeds with the one-to-one substitution of the N atoms by O atoms to form stable O$_\mathrm{N}$ sites without universal phase transition to $\beta$-Ga$_{2}$O$_{3}$ structures~\cite{Ma2022Exploring, Qi2021Bridging}, therefore, the atomic ratio of Ga/(N+O) is set to be 1 in the calculations. 

On the other hand,
in our experiments, the non-thermal oxidation caused by OPT is a far-from-equilibrium process, so the referenced reaction is not initiated from the stable chemically segregated reactants Ga+Ga$_{2}$O$_{3}$, but can be the chemically mixed states such as unrelaxed wurtzite GaO, strain-relaxed wurtzite GaO, and lattice distorted GaO. The exemplary supercells of GaO in these strain states can be found in Supplementary Information Figure S6.
Moreover, it can be clearly seen from the TEM image (Figure~\ref{fig:Fig1}b) that, although the lattice constant of the top GaON layer is different from that of the wurtzite GaN, the GaON layer still maintains a certain crystallinity.
As a result, the reactant that is most relevant to our experiments, as shown in Figure~\ref{fig:Fig3}a (ii), should be ``GaO (relaxed)", where the lattice strain is released, and the hexagonal wurtzite lattice is not heavily distorted.

As shown in Figure~\ref{fig:Fig3}b, the enthalpy, $H$, in eV per chemical formula unit (eV/f.u.) is plotted against the fraction of oxygen, $x$, in the formula unit set as GaO$_{x}$N$_{(1-x)}$. 
The colored lines label the combination of reactants (GaO at different strain states combined with GaN) without chemical reaction, %in Equation~\ref{eq:Eq2}
while the black data points are the calculated enthalpy of the ``reacted" GaO$_{x}$N$_{(1-x)}$ supercells (product) at different strain states. 
Each black data point is averaged over five independent supercells with the corresponding $x$. 
Three strain states of the ``reacted" GaO$_{x}$N$_{(1-x)}$  supercells (product) are considered: (a) static wurtzite lattice adopted to the lattice constants of perfect GaN with fixed-position O$_\mathrm{N}$ sites (unrelaxed); (b) static lattice with internally relaxed O$_\mathrm{N}$ sites (constraint); and (c) relaxed lattice in $z$-axis with fully relaxed O$_\mathrm{N}$ sites ($z$-axis relaxed). 
In this way, the energy differences attributed to the internal O$_\mathrm{N}$ relaxation and the external strain release can be clearly revealed.
It is shown that the internal relaxation of the O$_{N}$ sites results in a main reduction in enthalpy, while the external strain release becomes significant only at the HOC region (x $>$ 0.5). 
More essentially, the comparison of the $H$ between the products and the reference reactants gives the enthalpy of formation (heat of mixing) that governs the thermodynamic evolution of the system.

As the orange line is the most relevant condition in our experiments, we further plot the enthalpy of formation (heat of mixing), $\Delta H$, of this condition as shown in Figure~\ref{fig:Fig3}c. 
The positive and negative $\Delta H$ values indicate endothermic and exothermic reactions, respectively.
Intriguingly, a balanced point around $x$ = 0.33 is seen in Figure~\ref{fig:Fig3}c. 
During the annealing process, both the internal local site and external strain are gradually relaxed to more stable states, and the migration of the O atoms should follow the same trend.
Therefore, in a solid-state reaction with no external flux of the O atoms (as the annealing was conducted in N$_{2}$), the uniform system with a preset fraction of oxygen will tend to segregate into two HOC and LOC systems if the overall change in the $\Delta H$ is negative. 
Indeed, with reference to ``(1-x)GaN + (x)GaO (relaxed)", the energy landscape shows that the system (e.g., at $x$ = 0.5) is energetically favorable to finally segregate into pure GaN and GaO$_{0.9}$N$_{0.1}$ systems, so the HOC and LOC nanolayers seen in the experiments can be explained as an intermediate metastable state before the full segregation. 

The DFT calculated structures of the LOC-GaON are consistent with the TEM image (Figure~\ref{fig:Fig1}b). As mentioned in Figure~\ref{fig:Fig2}d, the $x$/$y$ in LOC-GaON of the 900 \degree C-annealed sample is 0.54 ($\pm$0.05)/1.
Based on this ratio, the possible crystal structures are calculated ($x$/$y$ of 0.5/1 is used in the DFT calculation for simplicity), as shown the Supplementary Information Figure S6, in which the LOC-GaON exhibits the original wurtzite structure. 
As a result, only the HOC-GaON with the heavily distorted lattice can be identified in the TEM image, while the LOC-GaON layer is not distinguishable. 
Therefore, as shown in Figure~\ref{fig:Fig3}d, we present a schematic illustration of the oxygen migration in GaN involving fast local substitution in $\mu$s time scale (based on the Arrhenius equation) and the slow overall segregation happening in minute time scale. 
The initial and final thickness of the GaON nanolayer is the same as the data in Figure~\ref{fig:Fig1}d. 

\begin{figure*}[ht!]
  \includegraphics[width=17cm]{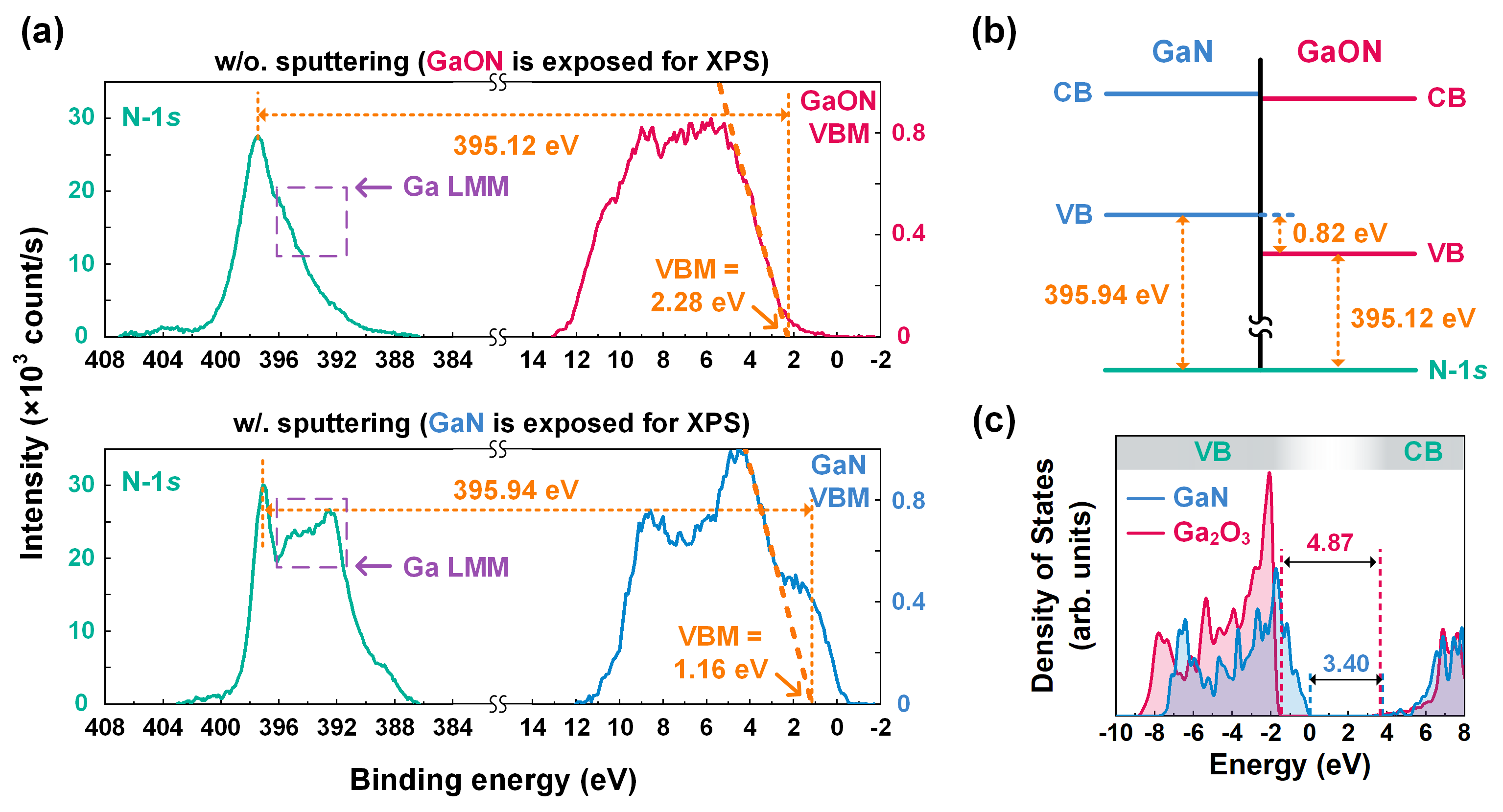}
  \caption{(a) The XPS spectra of the sample after OPT and 900 \degree C annealing, before (upper panel) and after (lower panel) Ar$^{+}$ sputtering.
  The peak of the N-1$s$ core level and the VB are used to determine the valence band offset between GaN and GaON in (b).
  The difference of the Ga LMM Auger intensities is owing to the formation of the surface metallic Ga after Ar$^{+}$ sputtering~\cite{huang2018ion}.  
  (b) Illustration of VB offset between GaN and GaON. 
  (c) The DOS of pure wurtzite GaN and Ga$_{2}$O$_{3}$ calculated using DFT.  
 }
  \label{fig:Fig4}
\end{figure*}

In practical applications, the energy band diagram is essential for understanding and predicting the performance of the devices. 
Thus, understanding the band alignment of the GaON nanolayer to the GaN substrate is critical to pave the way for its device-level applications. 
The \textit{in-situ} Ar$^{+}$ sputtering is used to remove all the GaON on the surface, so the GaN underneath is exposed for XPS characterization.
By comparing the XPS spectra of the valence band (VB) levels, the offset of the VB maxima (VBM) between GaON and GaN can be determined. 
As shown in Figure~\ref{fig:Fig4}a and b, using the N-1$s$ core level as a reference, the VBM offset is 0.82 eV, and the VBM of the GaON nanolayer is lower than that of the GaN substrate. 
It is noticed that the Ga LMM Auger peaks (purple region in Figure~\ref{fig:Fig4}a) and the additional bumped upper tail of GaN VB are originated from a thin layer of metallic Ga on the GaN surface introduced by \textit{in-situ} Ar$^{+}$ sputtering (Ga-Ga curve in Figure~\ref{fig:Fig2}b). 

As for the conduction band (CB) offset, because the CB minimum (CBM) is mainly contributed by the Ga-4$s$ orbital, it is expected that the CBM offset should be rather minimal. 
Indeed, as shown in Figure~\ref{fig:Fig4}c, the density of states (DOS) from the DFT calculation indicates that the CBM offset is $\sim$0.1 eV. 
Here, the comparison is made between the pure GaN and pure Ga$_{2}$O$_{3}$ systems, where the VBMs are contributed by occupied N-2$p$ for GaN and O-2$p$ for Ga$_{2}$O$_{3}$, respectively. 
Therefore, a larger VBM offset ($\sim$1.37 eV) is seen. 
In contrast, the marginal CBM offset can be understood with respect to other similar common-cation compound semiconductors in terms of simple chemical trends of the band edge positions and the hydrostatic volume deformation potential of the $\Gamma$ state~\cite{li2006ab, swallow2021indium}.
Specifically, in Ga-cation systems, the energies of the CBMs at $\Gamma$-point with insignificant shift indicate a small atomic volume deformation of Ga atom in different systems.

\subsection{Device-level demonstration: Diodes and $p$-FETs} \label{sec:dev}

Based on the experimental and computational results, the GaON nanolayer is highly compatible with GaN-based platform by nature and can be readily integrated into electronic-device applications.
Therefore, we further demonstrate its utility with Schottky junction diodes and GaN $p$-channel field effect transistors ($p$-FETs).

\begin{figure*}[ht!]
  \includegraphics[width=17cm]{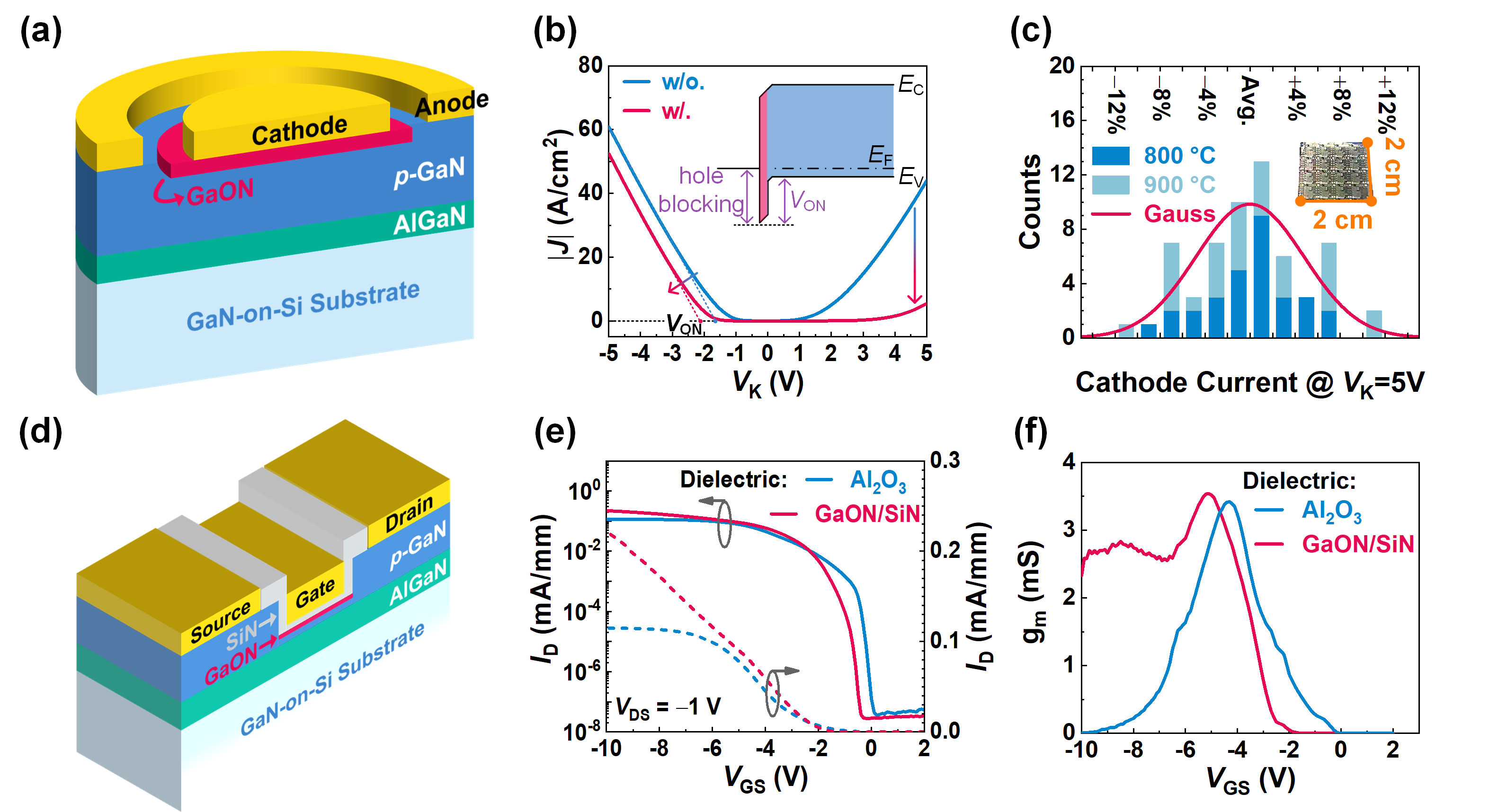}
  \caption{(a) Structure of fabricated GaON/GaN Schottky junction diode. 
  (b) Comparison of the current densities between diodes with and without GaON nanolayer. $V_\mathrm{K}$ stands for the cathode voltage. Inset figure is the energy band diagram of  GaON/GaN Schottky junction diode.
  %The solid and dashed lines are the same data plotted in log (left) and linear (right) scales, respectively. 
  (c) Statistic histogram of the cathode current at $V_\mathrm{K}$ of 5 V for 60 independent devices made on two 2 $\times$ 2 cm$^{2}$ samples (annealed at 800 and 900 \degree C, individually). 
  (d) Structure of fabricated GaN $p$-FETs. The gate dielectric in controlled group is Al$_{2}$O$_{3}$, while it is GaON/SiN$_{x}$ stack in experimental group. 
  (e) Transfer curves in log (left) and linear (right) scales of the GaN $p$-FETs with Al$_{2}$O$_{3}$ and GaON/SiN$_{x}$ gate dielectrics. 
  (f) Transconductance ($g_{m}$) of the GaN $p$-FETs with Al$_{2}$O$_{3}$ and GaON/SiN$_{x}$ gate dielectrics. The detailed fabrication process is included in Supplementary Information Figure S7 and S8.
 }
  \label{fig:Fig5}
\end{figure*}

Figure~\ref{fig:Fig5}a shows the schematic structure of the fabricated Schottky diode with the GaON nanolayer (GaON diodes), of which the fabrication process is described in the Supplementary Information. 
The conventional diode without the GaON nanolayer ($p$-GaN diode) has also been fabricated on the same epi-structure for comparison. 
Owing to the large metal/GaON barrier as shown in the inset figure of Figure~\ref{fig:Fig5}b, the GaON diodes exhibit an enhanced hole blocking capability with average two-order-of-magnitude smaller reverse-bias leakage current density compared with the conventional $p$-GaN diodes, and thus an enhanced current-regulating performance (Figure~\ref{fig:Fig5}b). 
Although the forward turn-on voltage ($V_{ON}$) of the diodes is increased (by $\sim$0.47 V from linear extrapolation) with the GaON being introduced, the differential on-resistance ($R_{ON}$) of the diodes at the ON-state is seldomly affected ($<$ 0.7$\%$) by the GaON nanolayer. 
The additional $V_{ON}$ assists holes to overcome the additional barrier induced by the GaON when the GaON diode is fully turned on, after which the GaON would not further limit the transportation of holes, so the differential $R_{ON}$ is not affected.
The increse on $V_{ON}$ is smaller than the VBM difference between GaON and GaN (0.47 eV versus 0.82 eV), probably due to the tunneling process, as the GaON nanlayer is only 3.99 nm in thickness.
The uniformity of the GaON nanolayer on a large scale is revealed by analyzing 60 independent GaON diodes fabricated on two 2 $\times$ 2 cm$^{2}$ samples (with OPT, and annealed at 800/900 \degree C, respectively). 
The statistical distribution of the reverse leakage current at $V_\mathrm{K}$ = 5 V is shown in Figure~\ref{fig:Fig5}c. 
A reasonable uniformity with a standard deviation within 4.9$\%$ has been found, verifying the potential of the proposed GaON nanolayer in large-scale fabrication.

Figure~\ref{fig:Fig5}d presents the schematic structure of the GaN $p$-FETs with a novel GaON/SiN$_{x}$ gate dielectric stack. 
For comparison, the GaON/SiN$_{x}$ dielectric stack of the devices is replaced with conventional Al$_{2}$O$_{3}$ in controlled samples. 
The device performance is summarized in Figure~\ref{fig:Fig5}e and f. 
The GaN $p$-FETs with GaON/SiN$_{x}$ dielectrics have larger peak transconductance ($g_{m}$) and increased maximum channel current ($I_{D}$). 
In conventional Al$_{2}$O$_{3}$-gate devices, the Al$_{2}$O$_{3}$ is used to blocking holes transporting from the channel to the gate, and hence reduces the gate leakage current. 
However, at the Al$_{2}$O$_{3}$/$p$-GaN interface, a large number of trap states would be charged by holes at high gate bias. 
The trapped holes would deplete part of the holes in the $p$-FET channel, resulting in the reduction of the channel current~\cite{ZZheng2021EDL}.
As a consequence, in the conventional Al$_{2}$O$_{3}$-gate device, the $g_{m}$ drops greatly with $V_{GS}$ $<$ -5 V.
To eliminate the influence of trap states in the gate region, a dielectric that has type-II alignement with $p$-GaN, such as SiN$_{x}$ in this demonstration, can be chosen. 
Such a type-II alignment allows the quick discharging of trapped holes through the VB of the SiN$_{x}$, as the trap states are within the VBM-offset energy range between the $p$-GaN and SiN$_{x}$. 
However, directly adopting SiN$_{x}$ will cause large gate leakage current, because there is no hole barrier from the channel to the gate.
This probblem can be well-solved by inserting a hole-blocking GaON nanolayer.
The GaON nanolayer has less trap states at its interface with $p$-GaN, because of its small lattice mismatch and similar symmetry against the hexagonal GaN.
Although trap states can appear at the GaON/SiN$_{x}$ interface, the trapped holes can be quickly discharged through the VB of SiN$_{x}$~\cite{zhang2021sin}.
As a result, the $g_{m}$ in the GaON/SiN$_{x}$-gate device maintains at reasonable values with $V_{GS}$ $<$ -5 V, benefiting to a higher maximum $I_{D}$.

In our previous works~\cite{ZZheng2021EDL, zhang2021sin, ZZheng2020EDL, zheng2021gallium}, we have discussed other advantages of this unique GaON nanolyer in boosting the performance of GaN $p$-FETs.
For example, the threshold voltage instability caused by interface trap states in conventional Al$_{2}$O$_{3}$-gate devices can also be greatly suppressed by this novel GaON/SiN$_{x}$ gate stack~\cite{ZZheng2021EDL, zhang2021sin}.
Meanwhile, in GaN $p$-FETs, as junction-less devices, gate-recess is an intuitive way to realize normally-OFF operation~\cite{Thomas2019EDLpFET}, while the etching-induced damages would decrease the hole mobility in the channel of $p$-FETs, as the channel is very close to the etched interface. 
An oxidized layer would help to deplete the holes under the gate region, realizing buried-channel normally-OFF $p$-FETs without aggressive gate-recess~\cite{ZZheng2020EDL, zheng2021gallium}. 

\subsection{Discussion and outlook: GaON nanolayer with multifunction}

So far, we have shown several superior structural and electronic properties of GaON nanolayer, such as enhanced thermal stability, large VB offset against GaN, and the two-step "oxidation-reconfiguration" process with highly tunable and compatible plasma techniques. 
Moreover, we have demonstrated that the GaON nanolayer can be readily integrated into GaN-based electronic devices to enhance the device performance. 
In this section, we will further discuss potential applications of the GaON nanolayer as a multifunctional structure on a GaN-based platform according to its attractive characteristics. 

\begin{figure*} [ht!]
  \includegraphics[width=17cm]{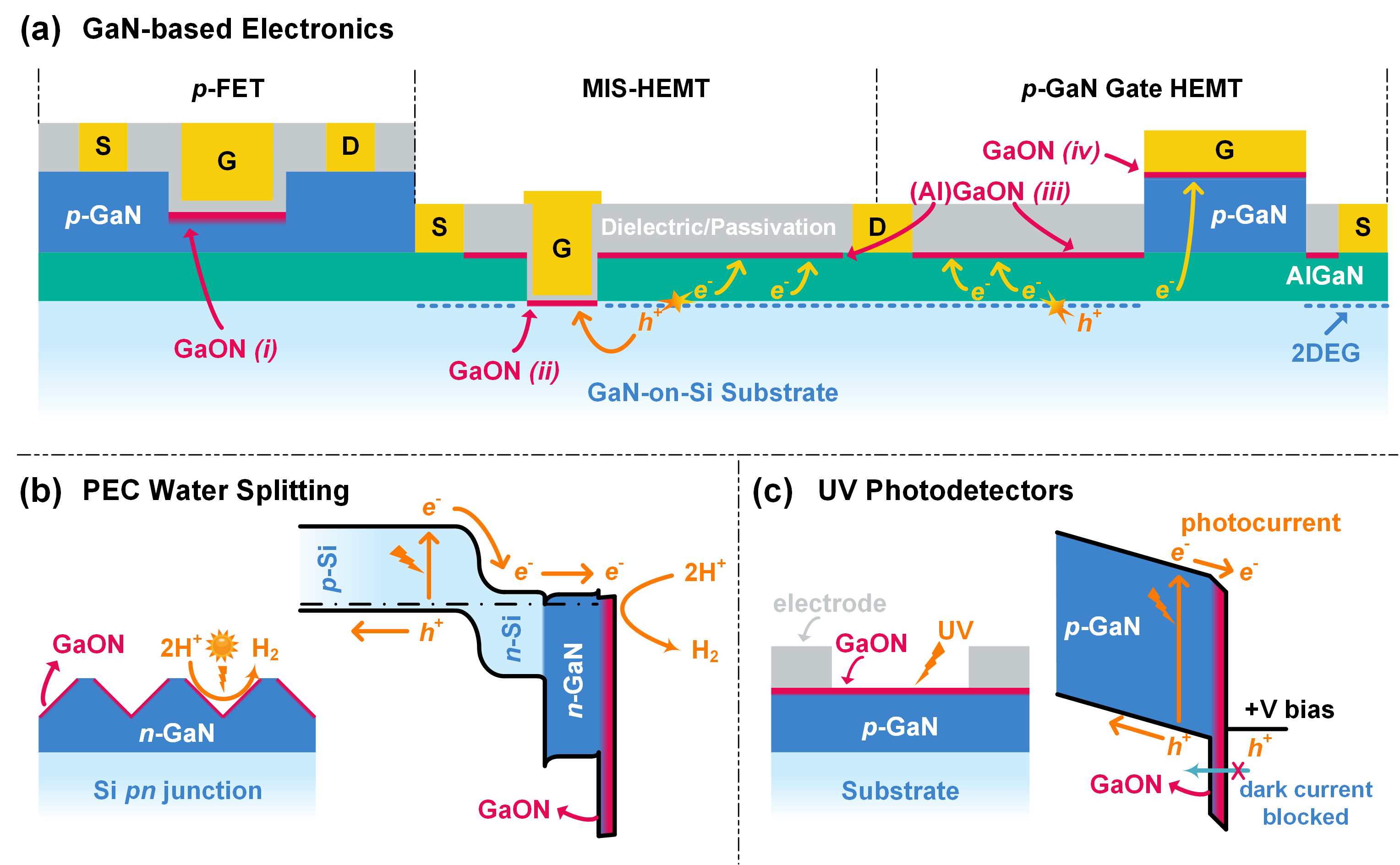}
  \caption{Schematic illustrations of the applicable scenarios of the GaON nanolayer: 
  (a) GaN-based HEMT for power electronics and $p$-FET for CMOS technology; 
  (b) passivation layer of photocathode in PEC water splitting;
  (c) dark-current-blocking layer in GaN-based UV photodetector.
 }
  \label{fig:Fig6}
\end{figure*}

The large VB offset against GaN makes GaON a naturally suitable dielectric for hole blocking. 
Compared with the conventional dielectric formed by deposition, GaON is \textit{in-situ} converted from GaN, featuring an uncontaminated high-quality interface. 
Moreover, its high thermal stability allows a large thermal budget process, such as low-pressure chemical vapor deposition (LPCVD)~\cite{MHua2018IEDM}, and metallization annealing. 

A straightforward application of GaON is as gate dielectrics in the GaN-based transistors.
Not only is it beneficial for the $p$-FETs (``GaON (i)" in Figure~\ref{fig:Fig6}a) as discussed in the previouse section,
%$p$-FETs. 
%GaN-based CMOS technology is essential for realizing high-density integration of gate drivers to GaN-based high electron mobility transistors (HEMTs)~\cite{RChu2016GaNCMOS}, thus further unlocking a high-frequency operation regime (e.g., $f >$ 1 MHz) for GaN-based power switching systems. 
%For the lagging-behind GaN $p$-FETs in the GaN CMOS system, in addition to the enhanced gate-control capability as discussed in Section~\ref{sec:dev}, the threshold voltage instability caused by interface trap states can also be greatly suppressed by adopting GaON as an interfacial layer in the MOS-gate stack (``GaON (i)" in Figure~\ref{fig:Fig6}a)~\cite{ZZheng2021EDL, zhang2021sin}.
%Meanwhile, in GaN $p$-FETs, as junction-less devices, gate-recess is an intuitive way to realize normally-OFF operation~\cite{Thomas2019EDLpFET}, while the etching-induced damages would decrease the hole mobility in the channel of $p$-FETs, as the channel is very close to the etched interface. 
%An oxidized layer would help to deplete the holes under the gate region, realizing buried-channel normally-OFF $p$-FETs without aggressive gate-recess~\cite{ZZheng2020EDL, zheng2021gallium}. 
%
%Not only is it beneficial for the $p$-FETs, this hole blocking capability of the 
the GaON nanolayer is also profitable in the $n$-channel GaN-based metal-insulator-semiconductor (MIS) HEMTs.
As shown in Figure~\ref{fig:Fig6}a, in a MIS-HEMT, the holes are mainly generated in the gate-to-drain access region from impact ionization, and these energetic holes will induce degradation of the dielectric by accelerated defects generation.
This hole-induced degradation can result in a highly concerned reliability issue that deducts the high-temperature reverse-bias (HTRB) operating lifetime of the MIS-HEMT~\cite{hua2018hole}.
However, the degradation can be suppressed by a hole-blocking GaON layer (``GaON (ii)" in Figure~\ref{fig:Fig6}a) at the interface between channel and dielectrics~\cite{MHua2018IEDM}. 

Another interesting feature of the GaON/GaN band alignment is the nearly flat CBM alignment despite a large VBM offset, remaining attractive in applications where require strong electron injection. 
As shown in Figure~\ref{fig:Fig6}b, an instance is as a protection layer in photoelectrochemical (PEC) water-splitting application, which can potentially overcome the stability-efficiency trade-off of Si photocathode~\cite{Scheuermann2016Design}. 
Because the Si surface is chemically unstable under water-splitting reaction, an essential protection layer, such as oxides and nitrides, is often adopted to protect the Si photocathode. 
However, employing an additional insulating layer will limit the carrier transportation from Si cathode to the water/solid interface and further result in additional voltage loss and saturation current degradation (efficiency loss)~\cite{Chen2011Atomic, Hu2014Amorphous, Scheuermann2016Design}. 
GaN recently rises much attention as the protection layer to Si photocathode owning to its chemically inert feature in water~\cite{Kibria2016Atomic}, and more essentially, nearly flat CBM alignment (offset of 0.16 eV) with Si, which ensures high efficiency of electron transportation from Si to the outer liquid-solid interface~\cite{Vanka2018High}. 
As the CBM alignment between GaN and GaON is nearly flat as well, so the potential efficiency loss during the electron transportation across the interface is rather eliminated. 
More intriguingly, GaON nanolayer on non-polar GaN facet has been found to be more active than GaN in terms of hydrogen evolution reaction (HER), because of the much lower free energy of hydrogen adsorption on the Ga site compared to that of pure GaN non-polar facet~\cite{Zeng2021Development}. 
Therefore, we note that it is highly promising that a large-scale GaON-nanolayer-decorated GaN layer (Figure~\ref{fig:Fig6}b) can be used to protect Si photocathode with enhanced performance of PEC water-splitting devices. 

One more application example is a GaN-based ultraviolet (UV) photodetector as shown in Figure~\ref{fig:Fig6}c.
The proposed GaON nanolayer can improve the device performance by suppressing the dark background current and thus enhancing the responsivity (sensitivity). 
Although GaN-based UV photodetector stimulates a lot of interest owing to its suitable detection range in UV, versatility and ability to serve in an extreme environment, a common challenge remains as the small photo-to-dark current ratio of less than 10$^{3}$~\cite{Abhishek2019Role, Goswami2021Fabrication}. 
Therefore, an oxide layer, such as ZnO and ZrO$_{2}$, is usually used to suppress the dark current that is mainly carried by holes in VB~\cite{Kim2020Investigation, Liu2015Suppression}. 
However, similar to the protection layer on Si photocathode, these oxides may undesirably reduce the photocurrent carried by photoelectrons if the interfacial CBM offset becomes large. 
Compared to other oxide materials, GaON nanolayer will not block the photon-generated electrons from the nearly-flat CBM alignment. 
Therefore, it can be used to improve the photo-dark current ratio by suppressing the dark current. 

Apart from the desired band alignment with GaN, the large bandgap of GaON also brings potential benefits, such as stronger electric-field strength and hot-carrier immunity. 
Therefore, the responsivity of GaN photodetectors can be greatly improved through a field-enhanced exciton ionization process triggered by a high electrical field~\cite{Kalra2018Demonstration, Tang2022Ga2O3}. 
By applying the GaON nanolayer, the device breakdown voltage can be boosted to a higher value~\cite{LZhang2021EDL}, allowing enough operating room for a high electrical field that enables the field-enhanced exciton ionization process. 
With such strong hot-carrier immunity, the GaON nanolayer is also suitable as a reinforcement layer to enhance the reliability of $p$-GaN gate HEMTs, as shown in Figure~\ref{fig:Fig6}a. 
For instance, the interface at AlGaN/passivation in the gate-to-drain access region (``(Al)GaON (iii)" in Figure~\ref{fig:Fig6}a) would undergo hot-electron bombardment during high-power switching operations. 
The bombardment would generate new surface states that capture electrons, resulting in a degradation of dynamic on-resistance ($R_{ON}$) and a degradation of reverse blocking capability. 
The AlGaN/passivation interface can be reinforced by an (Al)GaON nanolayer, and therefore, can withstand more hot-carrier bombardment~\cite{SYang2021EDL}.

Similarly, at the Schottky-type $p$-GaN gate region (“GaON (iv)” in Figure~\ref{fig:Fig6}a), the gate metal/$p$-GaN Schottky junction is also bombarded by hot electrons when a high gate voltage is applied, which will eventually lead to gate breakdown~\cite{TWu2015EDL}. 
A GaON nanolayer between the gate metal and $p$-GaN can significantly enhance the gate reliability and breakdown voltage, originated from the high immunity of the hot-carrier bombardment of GaON~\cite{LZhang2021EDL}.

As a final remark, we note that the potential applications of the GaON nanolayer are not limited to the discussed examples. 
Other recent studies on the GaON system, such as a smooth phase transition from GaN to $\beta$-Ga$_{2}$O$_{3}$ ~\cite{Ma2022Exploring} and deep oxygen incorporation in bulk GaN ~\cite{sun2006oxygen}, can extend the multifunctional utility of GaON in an unexpected way. 
For example, a possible GaN/Ga$_{2}$O$_{3}$ heterojunction with an intentionally embedded GaON nanolayer at the interface can make a high-quality transition layer in future GaN/Ga$_{2}$O$_{3}$-based devices.   

\section{Conclusion}

In this contribution, we proposed a well-controlled, large-scale, and yet few-step oxidation-reconfiguration approach to achieve metastable GaON epitaxial nanolayer on commercialized GaN platform. 
The construction as well as the formation mechanisms of the GaON nanolayer were clearly explained through well-established experimental and computational methods. 
The band alignment of the GaON nanolayer to the GaN substrate has been verified through experimental characterization and theoretical computation. 
The GaON nanolayer has been experimentally demonstrated to promote the performance of GaN-based Schottky diodes and $p$-FETs. 
Meanwhile, we discussed the employment of the GaON nanolayer as a multifunctional component in various kinds of applications, including GaN-based HEMTs for power electronics, $p$-FETs for CMOS technology, PEC water splitting, and photodetectors. 
This work can pave the way for future research on the GaON nanolayer as a multipurpose platform with inherent compatibility with a wide range of GaN-based cutting-edge applications.

% Experimental section
\section{Experimental and Calculation Section}

\subsection{Fabrication and characterization}
% The fabrication process of GaON nanolayer is summarized in Figure 1b.
The GaN wafers (from Enkris Semiconductor Inc.) were firstly cleaned with acetone and buffered-oxide-etchant to remove the organic contaminates and native oxide respectively.
% The atomic force microscopy (AFM) profiles (Figure S1, without oxygen plasma treatment (OPT) and no annealing) shows that the cleaned initial surface was of high quality with clear atomic steps and a low surface roughness of 0.588 nm.
In Step I, the samples were oxidized in an ICP chamber (Multiplex ICP from STS Inc.) with an O$_{2}$ flow of 30 sccm and platen/coil power of 20/50 W for 5 minutes.
% AFM profile confirmed that the surface was seldom affected by the low-power OPT (Figure 1a, with OPT and no annealing).
In Step II, the samples were annealed in N$_{2}$ at 800, 900, 1000 and 1100 \degree C, respectively for 55 minutes inside diffusion furnace (LB45 from ASM Inc.).
All samples were stored in acetone and were isolated from the ambient atmosphere. 
The bright-field TEM image was taken with JOEL (JEM-ARM200F) microscopy.  
% AFM profiles indicate that the samples started to decompose at 900 \degree C without OPT, but could withstand 1000 \degree C after OPT.
The further fabrication of GaON/$p$-GaN diodes is illustrated in Supplementary Information Figure S8.
The details of AFM, SIMS and XPS characterizations are given in Supplementary Information as well.

\subsection{Calculation details}
DFT calculations were conducted using the Vienna Ab-initio Simulation Package (VASP)~\cite{vasp1996}, employing the projected augmented-wave method~\cite{paw1994}. 
The electronic states were expended in plane-wave basis sets with an energy cutoff of 700 eV.
The Brillouin zone was sampled with a $\Gamma$-centered k-mesh grid with spacing 0.15 \r A$^{-1}$ which was equivalent to a dense 9$\times$9$\times$6 grid for a hexagonal wurtzite GaN unit cell.  
Gaussian smearing with a width of 0.03 eV was used to describe the partial occupancies of the electronic states. 
We chose 10$^{-7}$ eV and 5$\times$10$^{-3}$ eV/\r A as the energy and force convergence criteria for the optimization of the electronic and ionic structures, respectively. 
The Perdew-Burke-Ernzerhof (PBE) version of the generalized gradient approximation~\cite{pbe1996} was used for the initial configuration optimization and self-consistent energy calculations. 
The electronic densities of states were calculated using HSE06 hybrid functional with 32 $\%$ of the exact Hartree-Fock exchange~\cite{hse2006ori} to obtain correct bandgaps. More details is included in Supplementary Information.

\medskip
\textbf{Supplementary Information}  
\medskip

Supplementary Figs. S1–S8, Tables S1, Experimental and Computational Details and Discussions.

% Acknowledgements
\medskip
\textbf{Acknowledgements}
\medskip

This work was supported by the High-Level University Fund (G02236002 and G02236005) at the Southern University of Science and Technology (SUSTech). 
The computational resource was supported by the Center for Computational Science and Engineering at the Southern University of Science and Technology (SUSTech).
The authors acknowledge the Nanosystem Fabrication Facility (NFF) (CWB) of the Hong Kong University of Science and Technology (HKUST) for the device fabrication.
The TEM, SIMS and part of the XPS characterization of this research work was carried out in the Materials Characterization and Preparation Facility (MCPF) of the HKUST.

\medskip
\textbf{Contributions}
\medskip

M.H. and K.J.C. conceived the idea and supervised the project. 
J.C., S.F. and L.Z. fabricated the samples.
J.Z. performed the DFT calculation.
J.C., Y.C. and S.F. performed the TEM measurements and analysis.
J.C. and J.Z. performed the SIMS measurements and analysis.
J.C., J.Z., L.Z. and H.L. performed the XPS measurements and analysis.
J.C. performed the AFM measurements and analysis.
J.C. and J.Z. created the first draft of manuscript.
J.C., J.Z., Z.Z., X.C., Z.G., M.H. and K.J.C reviewed and edited the manuscript.
%\end{linenumbers}

\bibliographystyle{apsrev4-2} % give the final style for Nat. Mater.
\bibliography{final}% Produces the bibliography via BibTeX.
\end{document}